# USER REQUIREMENTS AND ANALYSIS OF PREECLAMPSIA DETECTION DONE THROUGH A SMART BRACELET

**Iuliana Marin[1], Andrei Vasilateanu[1], Bujor Pavaloiu[1], Nicolae Goga[1,2]**

[1]University POLITEHNICA of Bucharest, Faculty of Engineering in Foreign Languages (ROMANIA), marin.iulliana25@gmail.com
[2]Molecular Dynamics Group, University of Groningen, Groningen (NETHERLANDS), n.goga@rug.nl

## Abstract

Medical students along with the medical staff have to monitor the state of the patients by using modern devices which have to offer precise results in a short amount of time, so that the intervention to be done as soon as possible. They should be able to undergo different situations and to apply several theoretical cases for understanding the patient's state. These needs have been fulfilled during the last years due to the advance of Information and Communications Technology that allowed to register and interpret a large amount of data coming from patients who suffer from several illnesses. E-learning systems for blood pressure monitoring are used and new methods of patient observation, evaluation and treatment are applied compared to classical intervention. Doctors can analyze different cases and put a diagnosis through a collaborative virtual environment. Based on this, medical students can improve their knowledge for the practical training.

In the medical activities specialized devices occupy an important place. A device that can monitor the blood pressure is a smart bracelet that incorporates a pressure sensor along the wrist for continuous recording of blood pressure values. This enables the prediction of the emergency disorders using a decision support system. It facilitates the learning of new intervention approaches and boosts the responsiveness among learners. According to the World Health Organization, hypertensive disorders affect about 10% of pregnant women worldwide and are an important cause of disability and long-term death among mothers and children. This paper is based on a survey completed by persons of different ages and having various specialization domains regarding the use of smart bracelets for detecting preeclampsia. The aim is to decide upon its popularity among people and to determine the user requirements. The pregnant women will be constantly monitored, doctors can update the diagnosis of the patient, as well as they can discuss with each other through a collaborative virtual environment. Alerts will be sent to doctors in case of critical circumstances that threaten the life of the mother, as well as that of the baby. The medical students can learn from the critical situations and benefit from these cases while learning. The results of the survey showed that most of the interviewed persons consider the existence of such a device to be very useful, mostly the female individuals would feel more comfortable to have their blood pressure monitored during pregnancy.

## 1 INTRODUCTION

The latest tendencies in healthcare are focused on the patient's continuum monitoring and illness prevention. The patient's central role, mostly the one of the pregnant women, is illustrated in the preeclampsia detection done through a smart bracelet. The future of tomorrow depends on the actions of today.

Preeclampsia is an illness which characterizes pregnancies and it signifies the presence of hypertension and proteinuria due to which the life of the foetus, as well as the one of the mother are exposed to death [1]. It takes place after 20 weeks of gestation [2]. When proteinuria lacks, hypertension is accompanied by a low level of thrombocytes, impaired liver function, renal insufficiency, pulmonary edema, along with cerebral or visual disturbances [2].

When the woman gives birth the first time, the probability of preeclampsia occurrence varies between 10 and 14%, while for multiple pregnancies, the probability is between 2 and 7%, but 75% of the patients suffer from a beginning of preeclampsia [3].

In Romania the studies state that the annual incidence of preeclampsia ranges between 6 and 14%, respectively from 10-14% for the mothers who gave birth just once and 5.7-7.3% for those who gave birth multiple times [4]. The higher chance of preeclampsia occurrence is for the women who gave birth more than once, who suffered from preeclampsia before, but also for the women who gave birth once with the age under 20 and over 35 [4].

Blood pressure is measured in millimeters of mercury column (mm Hg). The larger number indicates systolic blood pressure, artery pressure, while the heart pumps blood. The lower number indicates diastolic blood pressure, pressure when the heart relaxes and fills with blood between beats [5]. For adults, normal blood pressure is below 120/80 mm Hg. High blood pressure (high blood pressure) is generally considered as a blood pressure greater than or equal to 140 mm Hg (systolic) or greater than or equal to 90 mm Hg (diastolic). Blood pressure analyzes in the pre-stress category (120-139 systolic or 80-89 diastolic) indicate an increased risk of high blood pressure. Hypotension is when a low blood pressure (below 90/60 mm Hg) is indicated.

People with heart disease, peripheral arterial disease, diabetes or kidney disease should have 130/80 or less. For pregnant women, hypertension is the most common medical problem encountered during pregnancy, complicating 5-10% of pregnancies. Mild preeclampsia is defined as the presence of hypertension in 2 situations, at least 6 hours apart, but without signs of internal organ damage, to a woman who was normotensive before 20 weeks of pregnancy [6]. Mean arterial pressure in the middle of the quarter has proven to be the best predictor of preeclampsia.

77% of the pregnant women who have preeclampsia do not know anything about this illness and therefore do not take measures [7]. It has been observed that the women which suffered due to preeclampsia are more exposed to cardiovascular diseases and diabetes of type II [8]. According to the existing statistics, black women are more likely to develop preeclampsia [9]. The presence of hypertension affects 5 up to 10% of the pregnancies. The probability of occurrence is low after the 37[th] week of pregnancy [1]. The increasing age of the mother boosts the health disorders occurrence [10].

The existent portable medical devices do not monitor the occurrence of preeclampsia and there is no such healthcare system which sustains this requirement. Along with this smart bracelet that monitors preeclampsia, the healthcare system embeds the doctor's reasoning and knowledge. Other doctors can also contribute and this external data will be verified and the inference engine will be updated.

By including Information Technology (IT) in healthcare is a necessity that will coordinate the clinical treatment of patients. Through the inclusion of smart device for daily use basis, the safety, accessibility and cost decrease are assured. The rate of appearance of this illness is increasing, being amongst the top maternal morbidity causes [11]. The average expenditure of this illness reaches the sum of €5243 per case in the European Union [12]. In United States, the costs caused by preeclampsia reach the sum of $2.18 billion per year [13].

According to the World Health Organization, hypertensive disorders affect about 10% of pregnant women worldwide and are an important cause of disability and long-term death among mothers and children [14]. Blood pressure affects around 40% of the world's population [15]. Early detection of abnormal blood pressure over time allows clinical monitoring and prompt therapeutic intervention. The unfortunate cases which are encountered in hospitals are most of the time caused by unmonitored, unknown pregnant women who suddenly reach the guard room of the hospital. In the world die each year 76,000 mothers and 500,000 infants [7].

In this paper we propose an innovative system for detecting preeclampsia. The proposed system will ensure accurate blood pressure monitoring, thereby increasing the likelihood of detecting early signs of blood pressure disturbance.

In the next section are presented some medical devices which measure blood pressure, along with information about the general architecture, actors, main functionalities, use cases and user requirements of the preeclampsia healthcare system. Section 3 outlines the importance of such a device based on a survey to which 100 persons answered. The last section summarizes the conclusions.

## 2 METHODOLOGY

The purpose of the research is to develop a system that will consist of a bracelet incorporating a pressure sensor along the wrist for continuous recording of blood pressure values. The resulting data will be sent via wireless connection to a smart phone and further to the server of the system. In this context, a software application will be developed that will predict the emergence of disorders using a decision support system.

Only some portable medical devices are available and commercialized. Among these are: H2 Care [16] made by a South Korean company that developed a bracelet located on the patient's wrist to measure blood pressure at 20 second intervals. The data is collected and subsequently managed by a system server. There is also an application on the mobile phone that can be set to remind users to repeat the measurement at fixed intervals.

HealthState [17], developed by a Singapore company, is a clock-like blood pressure monitor that allows you to measure blood pressure every 15 minutes within 24 hours.

A Canadian development company, LGTmedical, has continued to produce in mass a sensor and a mobile application, Phone Oximeter [18], which promises to improve the blood pressure measurement. This device uses a light sensor (which connects to the headphone jack of smartphones) to calculate the oxygen level in blood by measuring the absorption of two different wavelengths after passing through the finger. This solution is not comfortable because it is based on the self-control of the patient.

Traditional blood pressure monitors with inflatable cuffs are widely available for home use [19, 20]. The research intends to develop a portable device to monitor blood pressure and to predict hypertension during pregnancy.

A first innovative aspect is that a non-invasive sensor will be developed for continuous blood pressure monitoring that will come as an improved solution because the detection area will cover a large surface of the inner wrist compared to other blood pressure sensors on the market, where the sensor currently has a very small area and could report erroneous measurements due to non-alignment to the wrist artery. Therefore, compared to other competing solutions in the category of portable medical devices, this bracelet will have increased sensitivity and comfort, as the user will not have to worry about permanent contact in the artery area of the wrist for accurate blood pressure readings.

Compared to other solutions that cannot be worn, the benefits of the product are obvious and known: the user will not need to mount and dismantle the device repeatedly, will not take care of periodic blood pressure readings, while the portability of data measurement will facilitate the share with a doctor or medical staff.

The second innovative aspect of the product is the software application. It will continuously monitor the blood pressure throughout the pregnancy. The measurement of blood pressure is performed in the second and third trimesters to detect an abnormal blood pressure trend over time.

A predictive algorithm will be developed and implemented that will be able to report alerts much earlier to allow immediate intervention.

Given all of the above considerations, there is no similar solution on the market. The product will be accessible, technologically advanced, scalable and accessible. It is distinguished from rivals is made in terms of price, simplicity and functionality.

## 2.1 The actors of the preeclampsia monitoring healthcare system

The main users who will be implied are the monitored persons, respectively the pregnant women and the hypo/hypertensive persons for whom their blood pressure is monitored. Besides them, the doctors and the medical staff who treat the patient, together with the close relatives, caregivers and system administrators play a key role.

The monitored persons need to have their blood pressure continuously monitored. The data is collected by the bracelet using a software application, sent to a smart phone and then it reaches the application server. The close relatives, caregivers, doctors and the medical staff are the ones who receive alerts in case of a critical situation. The system administrator manages the healthcare system's configurations and user preferences.

## 2.2 Main functionalities of the preeclampsia healthcare system done through the bracelet

### 2.2.1 Data Transmission

The research aims to measure blood pressure. A sensor array will be placed on the artery in the wrist area. As the contact will not be perfect, the most difficult problems will be the processing and especially the calibration of the signals received for the blood pressure measurement. Another direction will be the data transmission between the intelligent bracelet and smart phone. At a interval of time, the measured blood pressure done by the bracelet is transmitted via the Bluetooth Low Energy (BLE) technology [21] which is often used for the Internet of Things [22] as well as for alerts if abnormal values are detected.

### 2.2.2 Detection of blood pressure diseases

The Android/iOS application will receive the data transmitted from the smart bracelet and interpret it. Research will be done to build models for the detection of the target blood pressure (hypo and hypertension, preeclampsia) based on real-time data and saved patient histories. The data will be secured and sent to a monitoring system server.

### 2.2.3 Clinical Decision Support

It will be a research to create a model and build a clinical decision support to assist medical consultations for diseases monitored by the bracelet system. It is expected to use a knowledge base and an engine of inference. The knowledge base includes information about the patient's blood pressure, history, and other relevant data, while the inference engine will apply logic rules to the knowledge base to aid the doctor's diagnosis and treatment process (see Fig. 1). This facilitates the learning of medical students and improved their response time in treating hypo and hypertensive illnesses.

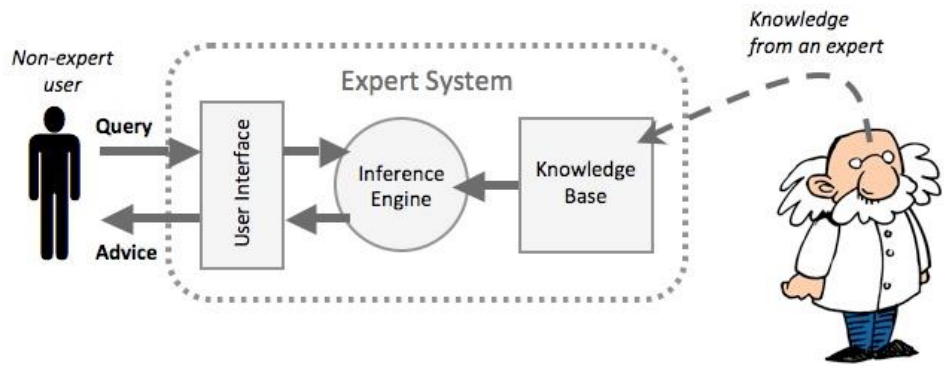

*Figure 1. Expert system [23]*

### 2.2.4 Web Application for doctors and medical staff

Some of the basic functionalities that will be created and the related research are as follows: patient history data will be extracted from the application's server where the database is located and will be decrypted. The patient's history will be viewable. The application will show suggestions for clinical decision support and will automatically provide a suggestion of diagnosis and treatment. The medical staff will be able to improve the diagnosis done automatically through annotations that will update the knowledge base on the server and will continue to be used to improve the intelligent support process to determine the treatment. This feature is very important, because the medical students will study new ways for treating their patients, as well as they will be aware of the possible results.

## 2.3 User requirements

The preeclampsia monitoring healthcare system has to assure the following user requirements:

### 2.3.1 Accessibility

The software application can be accessed from anywhere and from any operating system through the use of Internet. The computer which will access the application needs to have a processor with the frequency of at least 1 GHz and a RAM size greater than 1 GB.

### 2.3.2 Availability

Any software needs to be ready to be accessed immediately and in any moment. It is even more important in the present healthcare system. The downtime needs to be as low as possible, while maintaining an uptime of 99.999%.

### 2.3.3 Extensibility

The system needs to be able to incorporate new functionalities. In this way it can be measured the effort level needed to implement new features without having a major impact over the existent services. By including a web facility too, the system needs to be able to process more requests coming from the users in a short amount of time, therefore scalability is needed.

### 2.3.4 Performance

The system must have a short response time for completing a user request. Due to this, the execution time for any operation involving the use of the application needs to be reduced, as well as the transaction time for database operations, having a limited delay in the transfer of information.

### 2.3.5 Portability

The software application needs to operate in different environments, including operating systems, browsers, without changing the architecture. By having an application which is accessed using the Internet, it will work on any operating system.

### 2.3.6 Security

The client and administrator application will benefit from a secure authentication system, allowing the user to log into the system with dedicated credentials. Data encryption will be implemented for preventing intrusion and stealing of personal information.

## 2.4 System use cases

According to the actors which were presented in the Section 2.1 and the main functionalities of the preeclampsia healthcare system done through a smart bracelet, the use cases of the system are depicted in Fig. 2.

The system administrator manages the system configurations and the user preferences. The monitored persons will be able to have their blood pressure measured and interpreted, as well as to see details about their health state. The close relatives, caregivers, doctors and the medical staff will be able to see the monitored person's health state and they will get receive alerts in case of critical situations. The doctors will be to suggest, as well as to improve a diagnosis or a treatment.

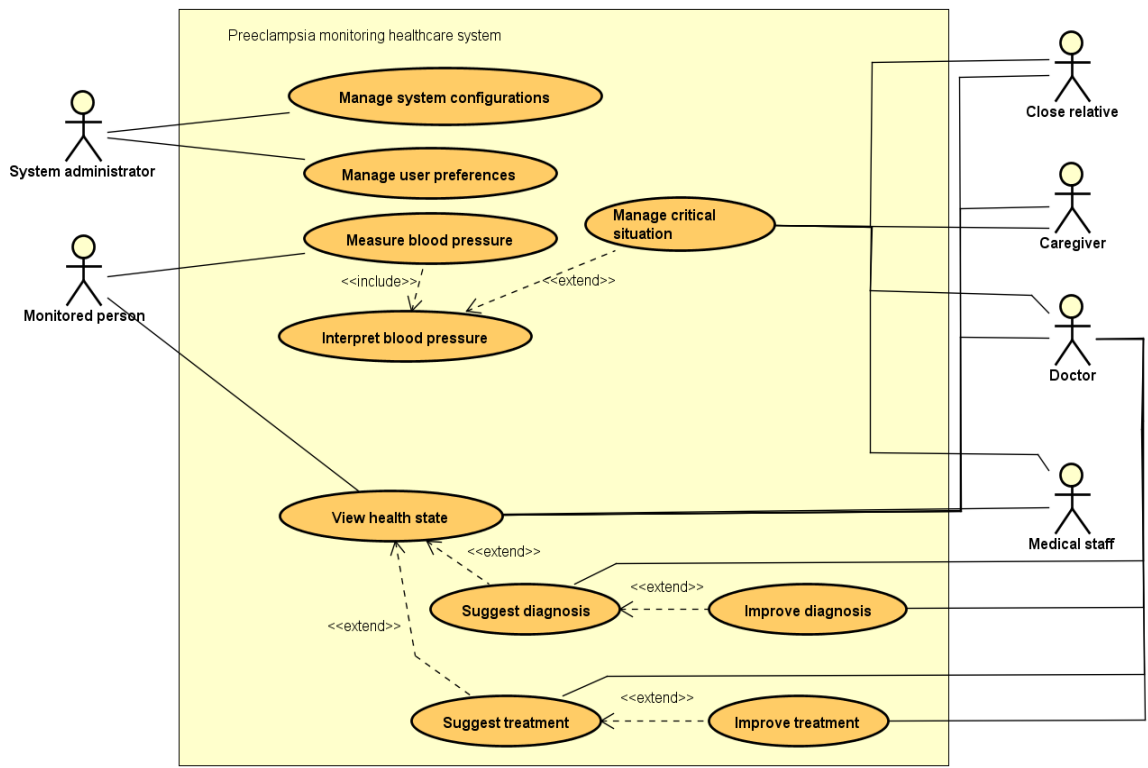

*Figure 2. Use case diagram*

## 2.5 General architecture of the system

A software application will control the operation of the system which is depicted in Fig. 3 and ensure data communication, acquisition and processing. The pressure sensor that detects the blood pressure waveform, which is a microfluidic sensor, will be placed on the wrist. The electronic module will provide signal conditioning and processing, and the data will be transmitted wirelessly to a smartphone. The clinical decision support will process the data and provide relevant physiological parameters that will be displayed numerically and graphically on an easy-to-use graphical interface. An SMS will be sent to the medical staff assigned to the patient when the blood pressure is below or above the normal level. Data is also transferred to the storage and processing system server.

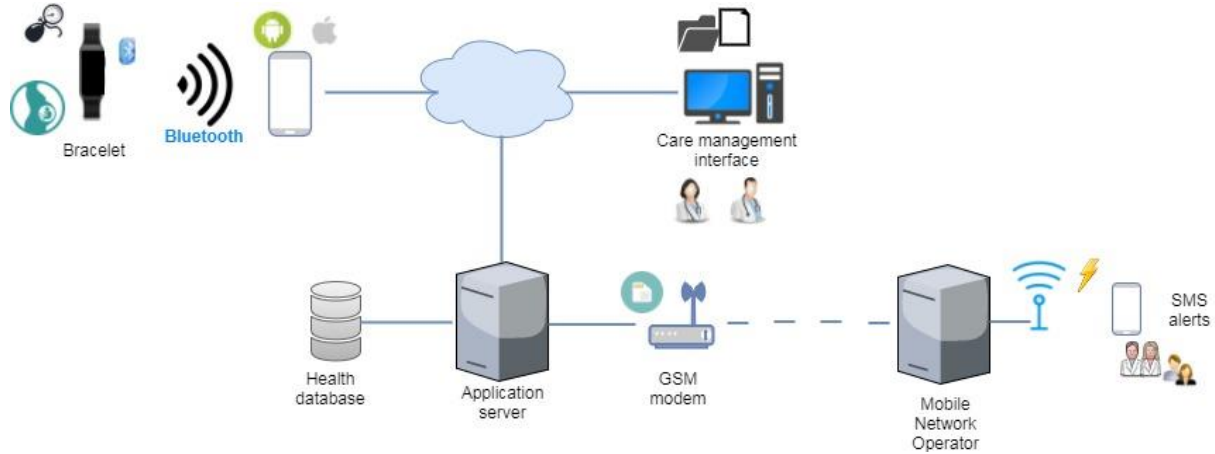

*Figure 3. Preeclampsia monitoring healthcare system architecture*

## 3 RESULTS

Based on the survey to which 100 persons answered, it is observed the necessity of such an electronic bracelet and the participants acknowledge the importance of monitoring the pregnant woman throughout her pregnancy. In the Section 2.2 have been mentioned the benefits which the system can offer for the medical students.

Out of the 100 participants, 64 were women and 36 were men, all having different ages, occupations and coming from different countries. Their average age is of 28 years. 71% of the interviewed persons own or intend to buy a smart watch or a smart bracelet. More than half of them already use a health care application on their smart device. 89% of the participants attested the fact that they would like the pregnant women to be monitored during gestation to determine the occurrence of preeclampsia. 12 persons out of the 100 suffer already from mild blood pressure, while one of them has a severe state of the illness. The most wanted features which the participants want are health statistics (71%), along with heart rate monitor (71%), followed by sports track (41%), message reminder (33%) and call reminder (24%).

85% of the interviewed persons trust the efficiency of such an intelligent bracelet and 89% of them said that they do not know any wearable device which monitors preeclampsia. 88.89% of them would recommend such a product to a relative or a friend, proving that such a product is welcomed and appreciated on the market.

### 3.1 Survey results from women

In Fig. 4 are illustrated the survey results from women, having an average age of 29 years. Most of them already own or intend to buy a smart watch or a smart bracelet. More than half already monitor their health using a health care application. An encouraging number of 90.63% participants are willing to have their blood pressure monitored during the pregnancy, 12.50% of them suffering already from mild blood pressure.

The features that they would like the bracelet to have are heart rate monitor (75%), health statistics (75%), followed by sports track (39.06%), message reminder (32.81%) and call reminder (20.31%). The cohort answered with a percentage of 95.31% that they do not know any wearable device which monitors preeclampsia. The same percentage of women which is willing to have their blood pressure monitored during pregnancy would also recommend such a product to a relative or a friend.

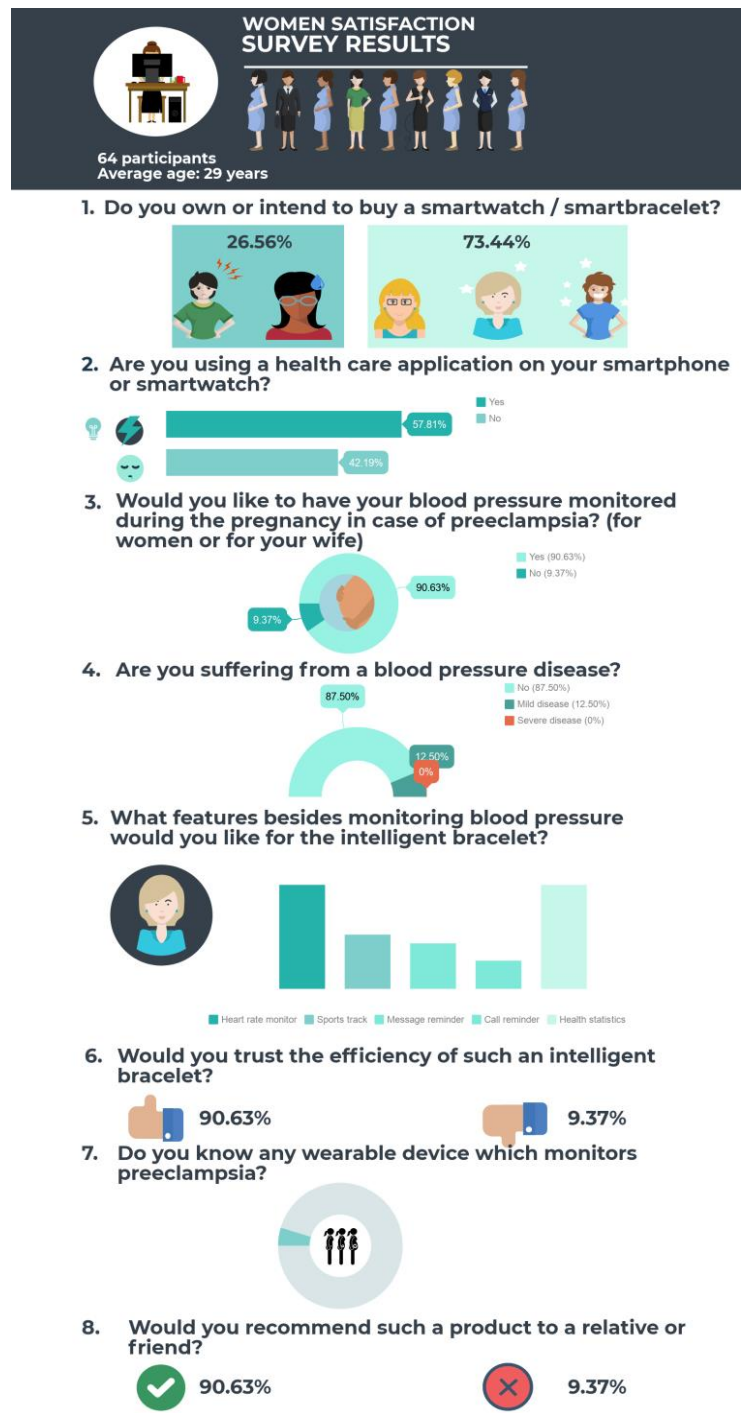

*Figure 4. Results from women*

## 3.2 Survey results from men

In Fig. 5 are illustrated the survey results from men, having an average age of 27 years. As like for the women participants, most of them already own or intend to buy a smart watch or a smart bracelet. Almost half of them already monitor their health using a health care application. 86.11% answered that they would like the women whom they know or their wife to have their blood pressure monitored during the pregnancy in case of preeclampsia. 11.11% of the men suffer from mild blood pressure, while 2.78% have a severe blood pressure illness.

The features that they would like to have available for the smart bracelet, besides monitoring blood pressure are health statistics (63.89%), heart rate monitor (61.11%), sports track (44.44%), message reminder (33.33%) and call reminder (30.56%).

Men are a little bit more sceptic and 75% of them would trust the efficiency of such an intelligent bracelet, but 85.71% would recommend this product to a relative or a friend. 77.78% of them answered that they not know any wearable device which monitors preeclampsia.

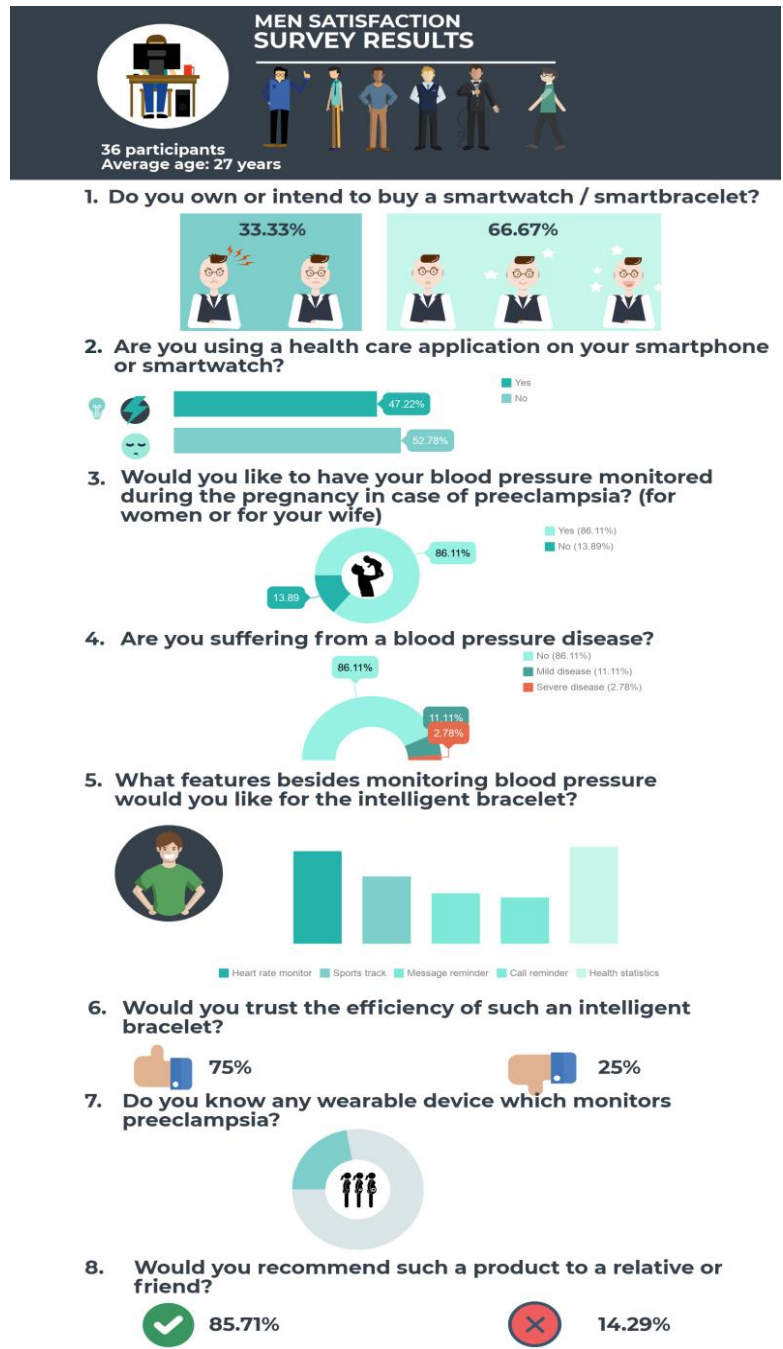

*Figure 5. Results from men*

## 4 CONCLUSIONS

An innovative approach in measuring blood pressure is done by incorporating a pressure sensor along the wrist for continuous recordings. Due to this the intervention is done in a shorter amount of time and it enhances the responsiveness among learners.

Out of the 64 interviewed women, 90.63% of them trust the efficiency of such an intelligent bracelet and they would recommend it to a relative or to a friend. The men share the same view as the women and they believe in the efficiency and importance of this device.

Besides the user requirements which result from the system description and the bracelet blood pressure monitoring feature, the other wanted characteristics are health statistics and heart rate monitor.

Doctors will be able to monitor the health state of the patients, as well as to assign and improve the diagnosis and treatment based on a collaborative virtual environment.

The need of preeclampsia prevention and monitoring has driven the development of such a healthcare system. Each year the percentage of women suffering from preeclampsia increases and that triggers the need of understanding, learning and developing knowledge about this illness, as well as hypo and hypertension from the current to the future medical specialists.